\documentclass[conference]{IEEEtran}
\IEEEoverridecommandlockouts
\usepackage{cite}
\usepackage{amsmath,amssymb,amsfonts}
\usepackage{algorithmic}
\usepackage[ruled,noline]{algorithm2e}
\usepackage{textcomp}
\usepackage{xcolor}
\usepackage{caption}
\usepackage{subcaption}
\usepackage{capt-of}
\usepackage{makeidx, epsfig, graphicx, setspace, longtable, multirow, float}
\usepackage{dblfloatfix}
\usepackage{cite}
\usepackage{color}
\usepackage[normalem]{ulem}
\usepackage{wrapfig}
\usepackage{hhline}
\usepackage{rotating}
\usepackage{subcaption}
\usepackage{array}
\usepackage{float}
\usepackage{booktabs}
\usepackage{hyperref}
\usepackage{comment}
\usepackage{xfrac}   

\def\BibTeX{{\rm B\kern-.05em{\sc i\kern-.025em b}\kern-.08em
    T\kern-.1667em\lower.7ex\hbox{E}\kern-.125emX}}

\def\Nats{\mathbb{N}}

\makeatletter
\renewcommand\section{\@startsection{section}{1}{\z@}%
                                  {-1.5ex \@plus -1ex \@minus -.2ex}%
                                  {0.8ex \@plus.2ex}%
                                  {\normalfont\normalsize\bfseries}}
\makeatother

\makeatletter
\renewcommand\subsection{\@startsection{subsection}{1}{\z@}%
                                  {-1.5ex \@plus -1ex \@minus -.2ex}%
                                  {0.8ex \@plus.2ex}%
                                  {\normalfont\normalsize\bfseries}}
\makeatother

\definecolor{codegreen}{rgb}{0,0.6,0}
\definecolor{codegray}{rgb}{0.3,0.3,0.3}
\definecolor{codepurple}{rgb}{0.58,0,0.82}
\definecolor{backcolour}{rgb}{0.91,0.91,0.91}

\newlength{\bibitemsep}
\newlength{\bibparskip}
\let\oldthebibliography\thebibliography
\renewcommand\thebibliography[1]{%
  \oldthebibliography{#1}%
  \setlength{\parskip}{\bibitemsep}%
  \setlength{\itemsep}{\bibparskip}%
}

\usepackage{setspace}

\usepackage[margin=1.009in]{geometry}
\graphicspath{{./}}

\newcommand{\etal}{\emph{et~al. \/}}
\newcommand{\eg}{\emph{e.g., \/}}

\usepackage{etoolbox,refcount}
\usepackage{multicol}

\newcounter{countitems}
\newcounter{nextitemizecount}
\newcommand{\setupcountitems}{%
  \stepcounter{nextitemizecount}%
  \setcounter{countitems}{0}%
  \preto\item{\stepcounter{countitems}}%
}
\makeatletter
\newcommand{\computecountitems}{%
  \edef\@currentlabel{\number\c@countitems}%
  \label{countitems@\number\numexpr\value{nextitemizecount}-1\relax}%
}
\newcommand{\nextitemizecount}{%
  \getrefnumber{countitems@\number\c@nextitemizecount}%
}
\newcommand{\previtemizecount}{%
  \getrefnumber{countitems@\number\numexpr\value{nextitemizecount}-1\relax}%
}
\makeatother    
{\end{itemize}%
\unskip\computecountitems\ifnumcomp{\previtemizecount}{>}{3}{\end{multicols}}{}}

\begin{document}

\title{\huge{Near Real-time Learning and Extraction of Attack Models from Intrusion Alerts}
}

\author{
Shanchieh Jay Yang, Ahmet Okutan, Gordon Werner, Shao-Hsuan Su, Ayush Goel, \& Nathan D. Cahill\\
\textit{Rochester Institute of Technology}\\
\textit{Rochester, NY, USA}
}

\maketitle

\thispagestyle{plain}
\pagestyle{plain}

\begin{abstract}
Critical and sophisticated cyberattacks often take multitudes of reconnaissance, exploitations, and obfuscation techniques to penetrate through well protected enterprise networks. The discovery and detection of attacks, though needing continuous efforts, is no longer sufficient. Security Operation Center (SOC) analysts are overwhelmed by the significant volume of intrusion alerts without being able to extract actionable intelligence. Recognizing this challenge, this paper describes the advances and findings through deploying ASSERT to process intrusion alerts from OmniSOC in collaboration with the Center for Applied Cybersecurity Research (CACR) at Indiana University. ASSERT utilizes information theoretic unsupervised learning to extract and update `attack models' in near real-time without expert knowledge. It consumes streaming intrusion alerts and generates a small number of statistical models for SOC analysts to comprehend ongoing and emerging attacks in a timely manner. This paper presents the architecture and key processes of ASSERT and discusses a few real-world attack models to highlight the use-cases that benefit SOC operations. The research team is developing a light-weight containerized ASSERT that will be shared through a public repository to help the community combat the overwhelming intrusion alerts.

\end{abstract}


\section{Introduction}

\emph{``There are just too many alerts. Can AI/ML summarize ongoing attack activities for us?'' -- Sentiment expressed by SOC analysts.}

\vspace*{4pt}

With more connected systems and applications, come new vulnerabilities and attack tactics at an ever-increasing rate. Cyber intrusion alert systems rightfully add sophistication and rule sets to catch up with new attack vectors. However, such complexity has also impeded interpretability and usability as Security Operation Center (SOC) analysts face a large volume and variety of noisy intrusion alerts. The overwhelming task of analyzing alerts often adds hours if not days of delay for action. Automatically summarizing intrusion alerts into interpretable attack patterns or behaviors, even if not perfectly capturing the truth, will enhance situational awareness and reduce time for threat hunting and incident analysis. 

To the best of our knowledge, there is no open-source projects, research, or commercial products that automatically summarize intrusion alerts into directly usable attack or threat models without excessive human intervention. There has been significant research on intrusion detection, \eg \cite{Lippmann2000, Kruegel2003, Livadas2006, Zhang2008, Li2012, Bilge2014, Blowers2014}, alert correlation, \eg \cite{Dain2002,Valeur2004,Ning2004,Qin2004,Yang2009,Ahmadinejad2011}, and attack graph generation, \eg \cite{Sheyner2002,Wang2006a,Ingols2006,Noel2009,Roschke2011,Fredj2015,Kaynar2016,Liu2019,Hu2020}. Alert correlation focuses on either grouping similar alerts or mapping alerts to a pre-defined attack template (probabilistic or rule based). Attack graph generation focuses on mapping system vulnerabilities to paths where attacker might take to penetrate the network. These works typically require significant expert knowledge or thorough network scanning to create templates or initial models. It is unfortunate that, even with the excellent research works, real-world SOC operations have not adopted them widely. Conversations with SOC analysts revealed many reasons behind the slow adoption of alert correlation and attack graph tools. One key reason is the complexity and expertise (and thus cost) needed to effectively use them in the real-world. A light-weight system that requires little expertise to configure,  adapts to changing attack behaviors, and provides intuitive summary of intrusion activities is needed. 

\begin{figure}[hbt]
    \centering
    \includegraphics[trim={0cm 0 0cm 0}, clip, width=0.45\textwidth]{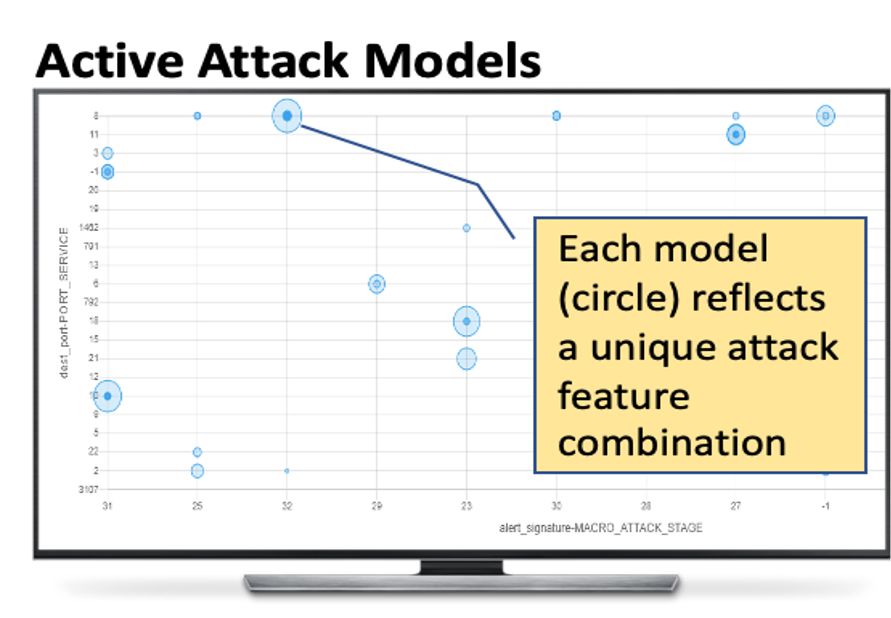}
    \vspace{2pt}
    \caption{A mock-up screen of ASSERT producing a number of attack models (each circle is a unique model).}
    \label{fig:screen}
\end{figure}

Recognizing this technology gap, this work builds upon prior works \cite{Assertv1, Assertv2} to enhance and deploy ASSERT to consume intrusion alerts collected through a real-world SOC operation - OmniSOC at Inidiana University. The resulting system is a light-weight information theoretic unsupervised learning system that consumes streaming alerts and synthesizes statistical attack models in near real-time without prior knowledge. An attack model is composed of intuitive characteristic features to describe a specific behavior or strategy, \eg a persistent arbitrary code execution and data exfiltration attack through VPN and SSH. The models are continuously updated to reflect evolving and emerging attack behaviors. The characteristic features provide the analysts with an interpretable summary for each unique attack behavior instead of fragmented evidences through the overwhelming noisy alerts. Figure \ref{fig:screen} shows a mock-up screen of ASSERT showing a number of attack models (each circle represents a unique attack model) by processing real-world alerts. We envision ASSERT feeding the attack models to existing commercial or open source SIEM platforms, \eg Elastic/Kibana, Splunk, OpenCTI, IBM X-Force Exchange, and FireEye iSIGHT, to ease the integration with SOC operations.

\vspace{-12pt}

ASSERT is a novel data driven solution to summarize the unknown mix of evolving attack behaviors even with rare or even unseen features. It accomplishes such in near real-time without requiring establishing initial models based on the system and network vulnerabilities. It is designed to help move SOC operations toward efficient and timely cyber defense. This paper presents the theoretical foundation and advances of ASSERT (Section \ref{sec:system}) and three example use-cases with real-world attack models to demonstrate its practical benefits (Section \ref{sec:models}). The concluding remarks are given in Sec \ref{sec:conclusion}.`

\section{ASSERT System Architecture and Advances}
\label{sec:system}

\begin{figure*}[hbt]
    \centering
    \includegraphics[trim={0cm 0 0cm 0}, clip, width=0.8\textwidth]{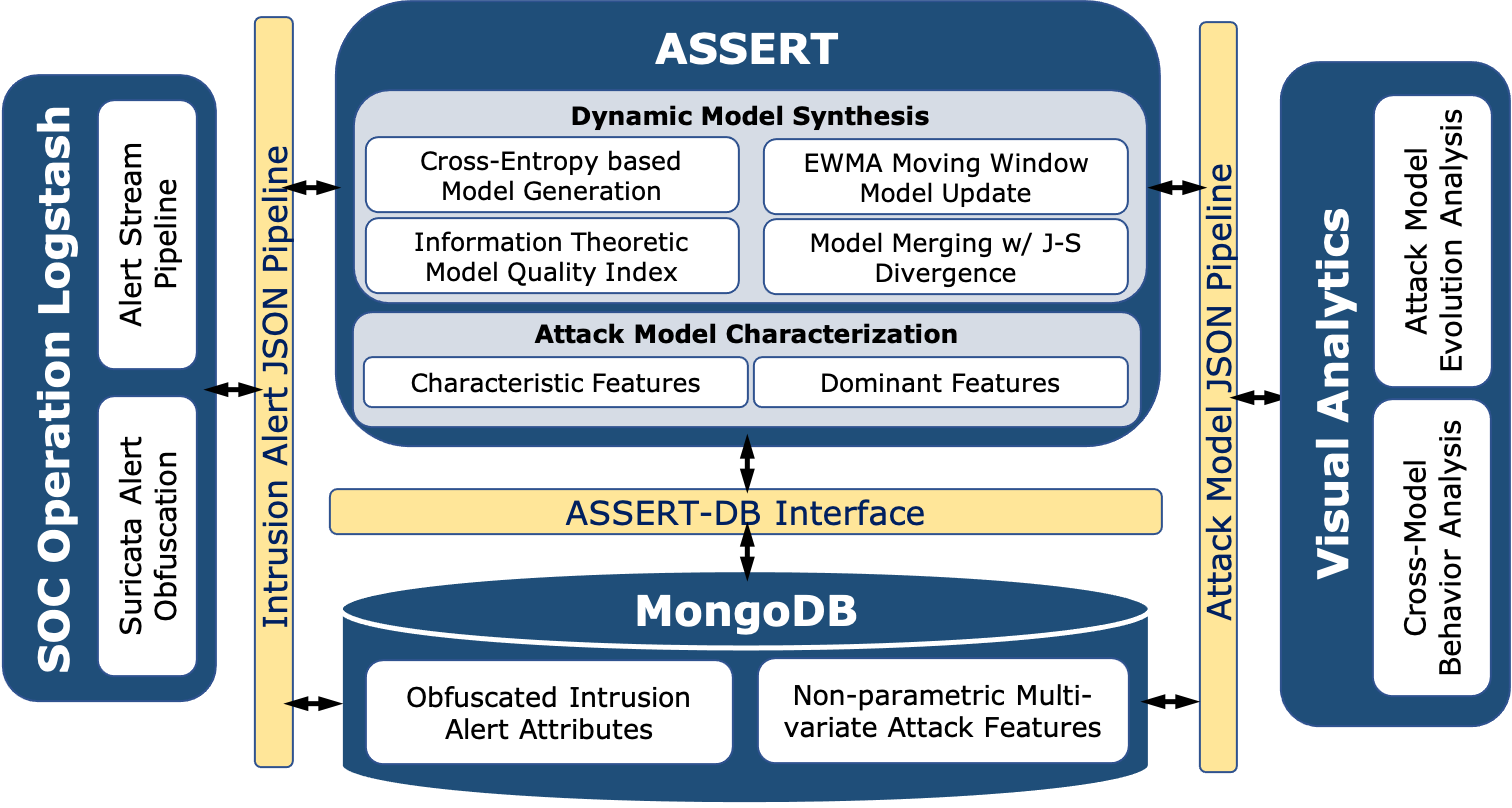}
    \vspace{4pt}
    \caption{ASSERT consumes alerts through logstash (left) and produces attack models for visualization (right).}
    \label{fig:arch}
\end{figure*}

Figure \ref{fig:arch} shows the system architecture of ASSERT, a significantly more efficient and deployable advancement comparing to the prior works \cite{Assertv1,Assertv2}. ASSERT takes in intrusion alerts through Logstash pipeline in JSON format, and produces attack models in JSON that can be used to produce visualization of the models' features and how they change over time. The following subsections describe ASSERT's key theoretical advances and development considerations.

\subsection{Theoretical Framework}
ASSERT takes in continuously arriving alerts triggered by malicious or incidental activities, transforms them into a homogeneous representation of attack actions $\mathcal{A}$, and dynamically creates and updates a set of models $\mathcal{Q}$, each representing a unique attack behavior. Let $\mathcal{A}=(\mathbf{a}^{(j)})_{j\in\Nats}$, where each action $\mathbf{a}^{\left(j\right)}$ has $m$ attack feature components $a_{i}^{\left(j\right)}$, $i = 1, \ldots, m$. Denote $\Theta_{i}$ to be the set of values that $a_{i}^{\left(j\right)}$ can take on; then $\mathbf{a}^{\left(j\right)}\in\Theta$, where $\Theta = \Theta_{1}\times\Theta_{2}\times\cdots\times\Theta_{m}$. We assume that all components take on a finite set of values; components such as temporal characteristics that might be real-valued can be quantized into a set of bins. Define $\mathcal{Q} = \left\{Q^{\left(1\right)}, \ldots, Q^{\left(K\right)}\right\}$ to be a set of random variables that represent different attack models. Note that the set $\mathcal{Q}$ can vary over time, since new models may be created, old ones may decay, and similar ones may be merged. Each $Q\in\mathcal{Q}$ is a random variable over $\Theta$ that is characterized by a (joint) probability mass function $p_{Q}:\Theta\rightarrow\mathbb{R}$ to reflect the collective attack behavior exhibited by a set of related $\mathbf{a}^{\left(j\right)}$'s. The attack features associated with model $Q$ are represented as a column vector $\mathbf{x}_Q\in\mathbb{R}^{N}$ via the mapping $\Gamma: \Theta\rightarrow\mathbb{R}^{N}$, $\Gamma(\mathcal{A}_Q) = \mathbf{x}_Q$, where $\mathcal{A}_Q$ is the set of $\mathbf{a}^{\left(j\right)}$'s to form the model $Q$. 
With the above framework, the key advances of ASSERT can be highlighted as:
\begin{enumerate}
\item {\bf Transform Streaming Alerts to Action Aggregates}: Design contextually meaningful and interpretable $\Theta$ and transform streaming alerts into homogeneous aggregates of attack actions $\mathcal{A}'\subseteq\mathcal{A}$ and $Q'$ with $p_{Q'}:\Theta\rightarrow\mathbb{R}$ and $\mathbf{x}_{Q'}:\Theta\rightarrow\mathbb{R}^{N}$ in near real-time.
\item {\bf Unsupervised Information Theoretic Synthesis of Attack Models}: Assess the `closeness' between $Q'$ and each $Q\in\mathcal{Q}$ and that between all $Q\in\mathcal{Q}$ to maintain an optimal $\mathcal{Q}$ composed of `unique' models, with near real-time update, creation, and merging processes.
\end{enumerate}

\subsection{Transform Streaming Alerts to Action Aggregates}
\label{sec:features}

Cyber intrusion alerts typically include time stamps, IP addresses, host names, port numbers, a description of the events, and other attributes specific to the monitoring tool or environment. ASSERT first transforms the streaming alerts produced by IDS's, \eg Suricata \cite{Suricata2020}, into aggregates of attack actions $\mathcal{A}'\subseteq\mathcal{A}$ over $\Theta$. Each aggregate represents a set of observed actions that are close in both temporal and spatial proximity. These aggregates reduce the system's sensitivity to individual alerts and serve as the basis to form the attack models.

\smallskip \noindent {\underline{\textbf{Homogeneous Multi-dimensional Action Space:}}}
The generic action space is designed to be meaningful to SOC analysts while allowing for near real-time computation. Without loss of generality, we propose to construct $\Theta$ to reflect the `how', `what', `where', and `when' of cyberattacks. Note that common practice recommends SOC operations to take advantage of `method intelligence' (how, where, and when) and `target/asset intelligence' (what), instead of spending excessive effort on often inaccurate and outdated `who' and `why'. Specifically, the ASSERT referenced in this paper utilizes the following action components with configurable weights to reflect the analyst's preferred emphasis between them.
\begin{itemize}
\item {\bf ($\Theta_\mathrm{a}$) Attack Intent Stages (AIS)} \cite{Moskal2020, Moskal2021}: a small set of categories built upon MITRE's ATT\&CK framework to reflect the plausible intended consequences of an attack action. This component will be discussed in more detail next.
\item {\bf ($\Theta_\mathrm{s}$) Targeted Services}: a mapping of port numbers to known services or labels indicating reserved or other uses of TCP or UDP ports.
\item {\bf ($\Theta_\mathrm{v}$) Attack Maneuver}: a mapping of IP addresses to categorical maneuvers, reflecting both the direction of each alert (inbound, outbound, internal) and the change(s) in source and destination IP addresses between consecutive alerts (\eg src\_is\_last\_dst and same\_src\_new\_dst) in the same `alert stream.'
\item {\bf ($\Theta_\mathrm{t}$) Time Elapsed}: the time elapsed since last alert in the same `alert stream.' The continuous measure in time are discretized in a logrithmic manner to reflect the significant variation (from nsecs to mins or hours) in attack speed. 
\end{itemize}
The definition of alert streams will be discussed after we present the details of AIS mapping. Additional attack action components are also available in ASSERT. 

\smallskip \noindent {\underline{\textbf{Inferring AIS ($\Theta_\mathrm{a}$) via Transfer/Active Learning:}}}
Descriptions of observed cyber events vary significantly from one IDS to another and can change over time to reflect new vulnerabilities or attack vectors. AIS \cite{Moskal2020} builds upon the MITRE ATT\&CK categories and enhances them to allow a generic and contextually rich interpretation of the intended consequences of observed actions regardless of the specific IDS. Manually mapping the large volume of alert descriptions ($\sim$64K Suricata alert descriptions) to AIS is an unrealistic task. We apply transfer learning to interpret the often cryptic alert descriptions by learning the cyber-specific terminologies and language structure through NVD and MITRE ATT\&CK framework. We investigated the tradeoffs in using different sizes and combinations of cyber postings and alert descriptions to train both the language model and the classifier. Note that transfer learning enables classifier training with only a relatively small number of alert descriptions (1$\sim$2\% of the total 64K Suricata alert descriptions). Once an alert-to-AIS model is built, it can be incrementally improved using active learning with pseudo labels \cite{Moskal2021}. The latest version of the alert-to-AIS model suggests 90\%+ accuracy in mapping the Suricata alerts to AIS. This mapping provides the homogenous $\Theta_\mathrm{a}$ to represent the observed attack actions.

\smallskip \noindent {\underline{\textbf{Alert Streams and Aggregation of Attack Actions:}}}
Finding relevant alerts is critical to derive $\Theta_\mathrm{v}$ and $\Theta_\mathrm{t}$ and to dynamically determine each aggregate $\mathcal{A}'\subseteq\mathcal{A}$. Intuitively, alerts are `relevant' if they occur closely in time and have common source(s) and/or target(s). Note that alerts that are close in time and have some common attributes do not necessarily mean they are absolutely executed by the same attacker. It simply means that there is a series of relevant actions within a contiguous time proximity and worthy of being analyzed collectively. Many alert aggregation efforts exist and shed lights on various contextually meaningful attributes, \eg \cite{Hofmann2011,Ahmed2014,Husak2017,Sun2020}. Different from many existing works, finding $\mathcal{A}'$ here is not meant to aggregate similar alerts into a meta-alert. On the contrary, ASSERT aims to find a set of alerts that are not identical but could potentially reveal symptoms of an attack episode. 

\begin{figure*}[hbt]
    \centering
    \includegraphics[trim={0cm 0 0cm 0}, clip, width=0.8\textwidth]{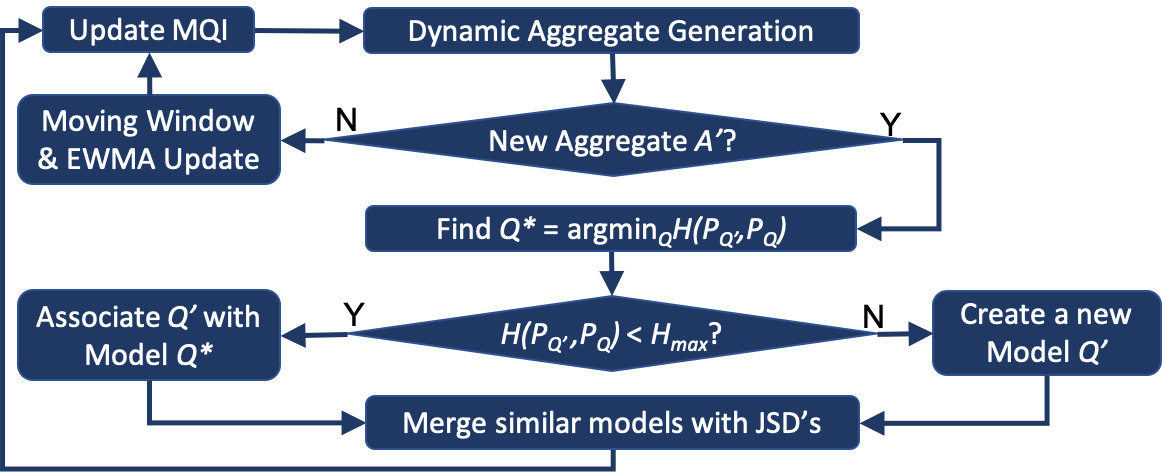}
    \vspace{4 pt}
    \caption{The high level processing flow of ASSERT.}
    \label{fig:assert_flow}
\end{figure*}

Aggregating alerts from a stream of alerts towards the same target means looking for the collective behavior from multiple sources (\eg DDoS), while doing so from the same source means seeking to reveal the specific attack behavior transitions (\eg from service discovery to targeted exploits). The ASSERT referenced in this paper considers alerts in the same stream if and only if they 1) originate from the same external source, 2) feed to that source, or 3) continue the maneuver (pivoting) internally in the targeted network. With this alert stream definition, ASSERT considers three options to find aggregates, or sub-sequences within each stream. First, a naive method considers consecutive alerts in the same aggregate if the time elapsed between them is smaller than a threshold. The second option applies a Gaussian filter to smooth the volume of alerts over time \cite{Moskal2018a} and extract attack episodes as aggregates. The third option utilizes control charts \cite{Raza2015} and Kolmogorov-Smirnov Test \cite{Massey1951} to dynamically learn and extract alert aggregates \cite{Werner2021}. For the use-cases shown in this paper, the first option was used.

\subsection{Unsupervised Information Theoretic Synthesis of Attack Models}
\label{sec:assert}

With the dynamically arriving $\mathcal{A}'\subseteq\mathcal{A}$ and the corresponding $Q'$, ASSERT updates $\mathcal{Q}$ in an unsupervised manner and near real-time. Figure \ref{fig:assert_flow} gives a high-level process flow on how ASSERT analyzes $\mathcal{A}'$ to dynamically update $\mathcal{Q}$. As the system processes continuous streaming alerts, it will update each model $Q$ with its $p_Q$ and $\mathbf{x}_Q$ to reflect changes in attack behavior. We use Exponentially Weighted Moving Average (EWMA) to update the effective $p_Q$ and $\mathbf{x}_Q$ with a moving window upon completion of processing each $\mathcal{A}'$. 

\smallskip \noindent {\underline{\textbf{Model Creation and Update with Cross-Entropy:}}}
Recall $Q'$ is a multivariate random variable with probability mass function $p_{Q'}:\Theta\rightarrow\mathbb{R}$. ASSERT needs to efficiently compare $p_{Q'}$ with $p_Q$ for all $Q\in\mathcal{Q}$ so as to determine whether to create a new model or update an existing one. The prior works \cite{Assertv1,Assertv2} find the max-posterior model under the Bayesian framework with entropy redistribution. The enhanced ASSERT finds
\begin{equation}
Q^* = \arg \min_Q (H(p_{Q'}, p_Q)) \enspace ,
\label{eq:crossentropy}
\end{equation}
where $H(p_{Q'}, p_Q)$ is the cross entropy between $p_{Q'}$ and $p_Q$. The cross entropy serves as a proxy for Kullback–Leibler Divergence (KLD) since $D_{\mathrm{KL}}(p_\mathcal{Q'} || p_Q)=H(p_{Q'}, p_Q)-H(p_{Q'})$, and $H(p_{Q'})$ is the same for all $Q\in\mathcal{Q}.$ This approach ensures adding minimal total information to represent the data if $Q'$ is associated with an existing model. It is also extremely fast since in practice $p_{Q'}$ typically contains only a few non-zero components in the summation when calculating the cross-entropy. 

With $Q^*$ found, the prior works \cite{Assertv1,Assertv2} explicitly compare the effects of adding $Q'$ to $Q^*$ versus creating a new model. While such process runs in polynomial time, it still takes significant memory and computation time. The enhanced ASSERT significantly simplifies the process by using a close form criterion: associate $Q'$ with $Q^*$ iff
\begin{equation}
H(p_{Q'}, p_Q) < \log\frac{|\Theta|}{\gamma} \enspace ,
\label{eq:crossentropy_cond}
\end{equation}
where $0\leq\gamma\leq 1$ is a multiplier to relax the restriction. The smaller the $\gamma$ is, the easier it is to associate $Q'$ with $Q^*$. Intuitively, with $\gamma=1.0$, the above condition corresponds to the case where the non-zero elements in $p_{Q'}$ are uniformly distributed in $p_{Q^*}$ with a total mass of $\sum_{j: q'_{j}\neq 0}q^{*}_{j}$. In other words, it is not desirable to associate $p_{Q'}$ with $Q^*$ since the non-zero elements in the new aggregate are no better than uniformly distributed with partial mass in $p_{Q^*}$. 

\smallskip \noindent {\underline{\textbf{Model Merging w/ Jensen-Shannon Divergence:}}}
The model synthesis process requires a way to assess whether the updated models over time become similar and need to be merged. Continuing the information theoretic approach, ASSERT adopts Jensen-Shannon Divergence (JSD) \cite{Lin1991}. The JSD metric is a symmetrical measure of the divergence between two models:
\begin{align}
&D_{\mathrm{JS}}(p_{Q_i} || p_{Q_j}) = \\
&\quad\quad\quad\quad \tfrac{1}{2} \left( D_{\mathrm{KL}}(p_{Q_i} || p_{\mathrm{avg}}) + D_{\mathrm{KL}}(p_{Q_j} || p_{\mathrm{avg}}) \right) \enspace , \nonumber
\end{align}
where $p_{\mathrm{avg}}=0.5(p_{Q_i}+p_{Q_j})$ represent the average of the two models. ASSERT uses an iterative heuristic shown in Algorithm \ref{alg:merging} to merge models when there exist pair-wise JSD smaller than a threshold. Complemeting the cross-entropy based model creation process, the merging heuristic works well in practice and typically takes fewer than 3 iterations to converge when invoked.

\begin{algorithm}
\DontPrintSemicolon
\SetAlgoNoLine
 $\mathcal{Q}$: The set of attack models;\\
 $D(i,j)$: JSD between $Q_i$ and $Q_j$;\\
 $D_\mathrm{max}$ = $\max_{i \neq j}(D(i,j)),~\forall Q_i, Q_j \in \mathcal{Q}$;\\
 $\delta$: Threshold for merging two models;\\
 $Q_\mathrm{l}$ = larger model that gives $D_\mathrm{max}$;\\
 $Q_\mathrm{s}$ = smaller model that gives $D_\mathrm{max}$;\\
 \BlankLine
 \While{$D_\mathrm{max} < \delta$}{
 $Q_\mathrm{l}$ = merge$(Q_\mathrm{l},Q_\mathrm{s})$;\\
 Update $\mathcal{Q}$;\\
 $D_\mathrm{max}$ = $\max_{i \neq j}(D(i,j)),~\forall Q_i, Q_j \in \mathcal{Q}$;\\
 }
\caption{JSD based Model Merging}
\label{alg:merging}
\end{algorithm}

\begin{figure*}[hbt]
    \centering
    \includegraphics[trim={0cm 0 0cm 0}, clip, width=0.98\textwidth]{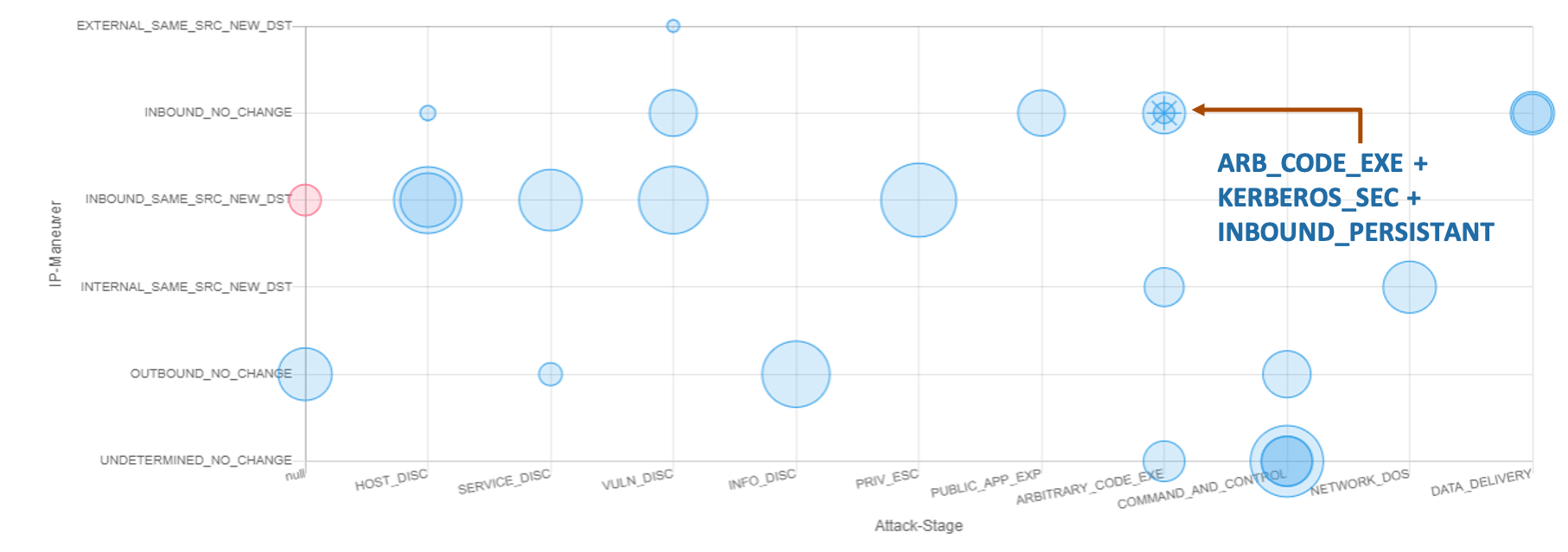}
    \caption{A screenshot of attack models produced by ASSERT. The model pointed by the arrow has characteristics features in Arbitrary Code Execution, Kerberos, and Inbound Persistent.}
    \label{fig:model1}
\end{figure*}
\subsection{Interpretation of Attack Models}
Each attack model $Q\in\mathcal{Q}$ is summarized with a set of characteristic features in $\mathbf{x}_Q$. For each action component, the characteristic feature is defined as follows:
\begin{equation}
x^*_Q = \arg \max_{x|\tilde{Q}\neq Q}(p_Q(x) \log p_{\tilde{Q}}(x)) \enspace ,
\end{equation}
where $p_Q(x)$ and $p_{\tilde{Q}}(x)$ are the probabilities of feature $x$ in $Q$ and $\tilde{Q}$, respectively. Intuitively, this finds the feature that is prominent (not necessarily the most) in $Q$ but very rarely or non-existence in any other model $\tilde{Q}$. The characteristic features provide an intuitive way for the analysts to comprehend the attack models. In practice, analysts quickly focus on the attack models that contains characteristic features they deem critical (\eg command and control or domain controllers).

\section{Case Studies}
\label{sec:models}
\begin{figure*}[hbt]
    \centering
    \includegraphics[trim={0cm 0cm 0cm 0}, clip, width=0.8\textwidth]{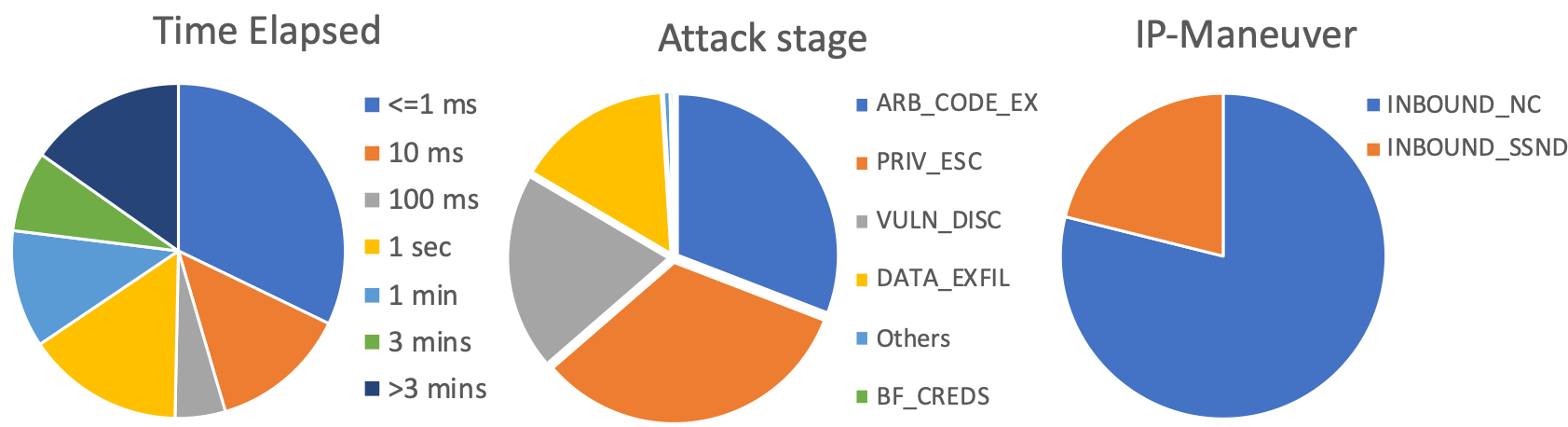}
    \vspace*{4pt}
	\caption{The detailed feature distributions of the Kerberos attack model.}
    \label{fig:model1-pie}
\end{figure*}

The ASSERT system was deployed to consume Suricata alerts generated from OmniSOC at Indiana University. On average, ASSERT received approximately 25K alerts per hour from OmniSOC while able to process 100K+ alerts per hour (up from $\sim$20K alerts/hr from the previous version) with a 4-core 32-GB memory Intel Xeon system. Normally, ASSERT will maintain about 20 to 25 models with the default configuration: 6-hour EWMA, $\gamma=2/3$ and feature weights of $(w_\mathrm{a}, w_\mathrm{s}, w_\mathrm{v}, w_\mathrm{t})=(0.3,0.3,0.3,0.1)$ for $\Theta_\mathrm{a}$, $\Theta_\mathrm{s}$, $\Theta_\mathrm{v}$, and $\Theta_\mathrm{t}$\footnote{Small variations of the configuration were used when deployed and experimented with OmniSOC data over the period of 3+ months, and thus for the case studies shown in the paper.}. Among the 20 to 25 models, most usually reflect the constant reconnaissance activities or false positives due to legitimate running services. The remaining few reflect suspicious attack behaviors that attract analysts' attention for further investigation. Note that the ASSERT system is not meant to determine whether an attack model is truly malicious; it is meant to reduce the needs for the analysts to sort through the overwhelming intrusion alerts by bringing their focus to the few critical attack models. We present a few interesting attack models below to demonstrate the use cases of ASSERT.


\subsection{Persistent Attack via Kerberos}
\label{sec:kerberos}
Kerberos is a network authentication protocol and susceptible to a number of vulnerabilities. Around late July 2020, OmniSOC saw about 300+ alerts over a period of time originated from the same sources that attempted various vulnerabilities of Kerberos. Figures \ref{fig:model1} and \ref{fig:model1-pie} shows a screenshot of the attack models produced by ASSERT with the arrow pointing to the attack model that characterizes the Kerberos attack and the feature distributions of the attack model (except the service which is entirely Kerberos), respectively. As can be seen on the Attack Stage feature pie chart (middle), there are several malicious activities, ranging from vulnerability discovery, privilege escalation, data exfiltration, to arbitrary code execution. Note that the mapping from Suricata signatures to Attack Stages is not perfect; however, the beauty of ASSERT is that, even with imperfect mapping, the collective behavior was clear to suggest a series of suspicious activities attacking through Kerberos. The 300+ alerts would be buried in the overwhelming 25K+ per hour alerts and likely became noise and be overlooked. ASSERT was able to extract this attack model out with its characteristic features in a timely manner.

\begin{figure*}[hbt]
    \centering
    \includegraphics[trim={0cm 0 0cm 0}, clip, width=0.85\textwidth]{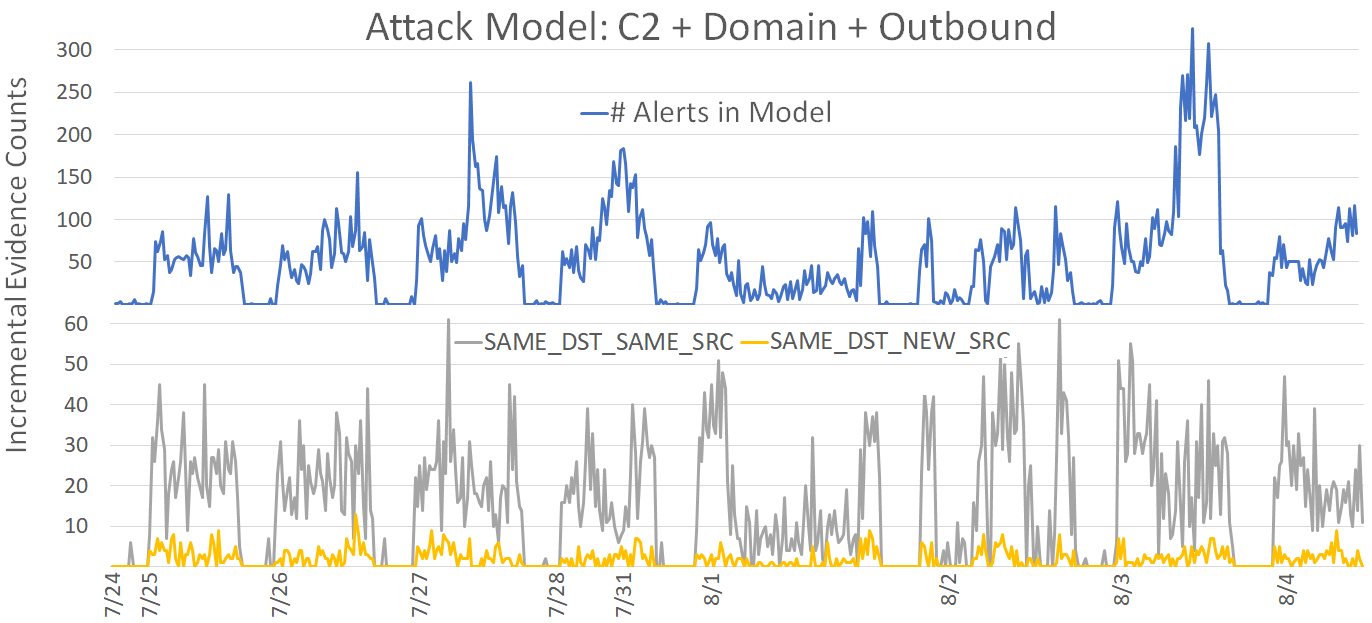}
    \caption{Alert volume tracked between July 24, 2020 and August 4, 2020 for Outbound C2 activities through domain controllers. The top figure shows the total alert volume and the bottom one shows the volumes for those originated from the same source or a new source between consecutive alerts (with the same external destination).}
    \label{fig:model2}
\end{figure*}

\subsection{Outbound C2 Activities}
\label{sec:c2+domain}
Over a period of approximately 11 days between July 24, 2020 and August 4, 2020, ASSERT tracked an attack model that reflected outbound activities resembling command-and-control (C2) attacks through Domain Controller related services. Figure \ref{fig:model2} shows the volume of alerts for this model over time (top) and the subset of alerts that have the same destination IPs but either the same source IPs or different source IPs between consecutive alerts (bottom). As can be seen, there are some semi-periodicity in these outbound C2 activities through domain controllers. The semi-periodical rising of total alert volume seems to coincide well with the alerts specifically originated from the same internal source. It is possible that there was legitimate traffic being flagged as C2 from the internal source. SOC analysts could take this attack model and analyze and compare specifically the source IP addresses (which the authors do not have access to) within the model. A plausible outcome of this analysis could be two-fold: 1) improve the system configurations to reduce flagging of legitimate traffic and 2) isolate the actual malicious C2 attacks through domain controllers. ASSERT was able to effectively extract, track, and present the critical features of suspicious outbound C2 activities through domain controllers continuously.

\begin{figure*}[hbt]
    \centering
    \includegraphics[trim={0cm 0 0cm 0}, clip, width=0.95\textwidth]{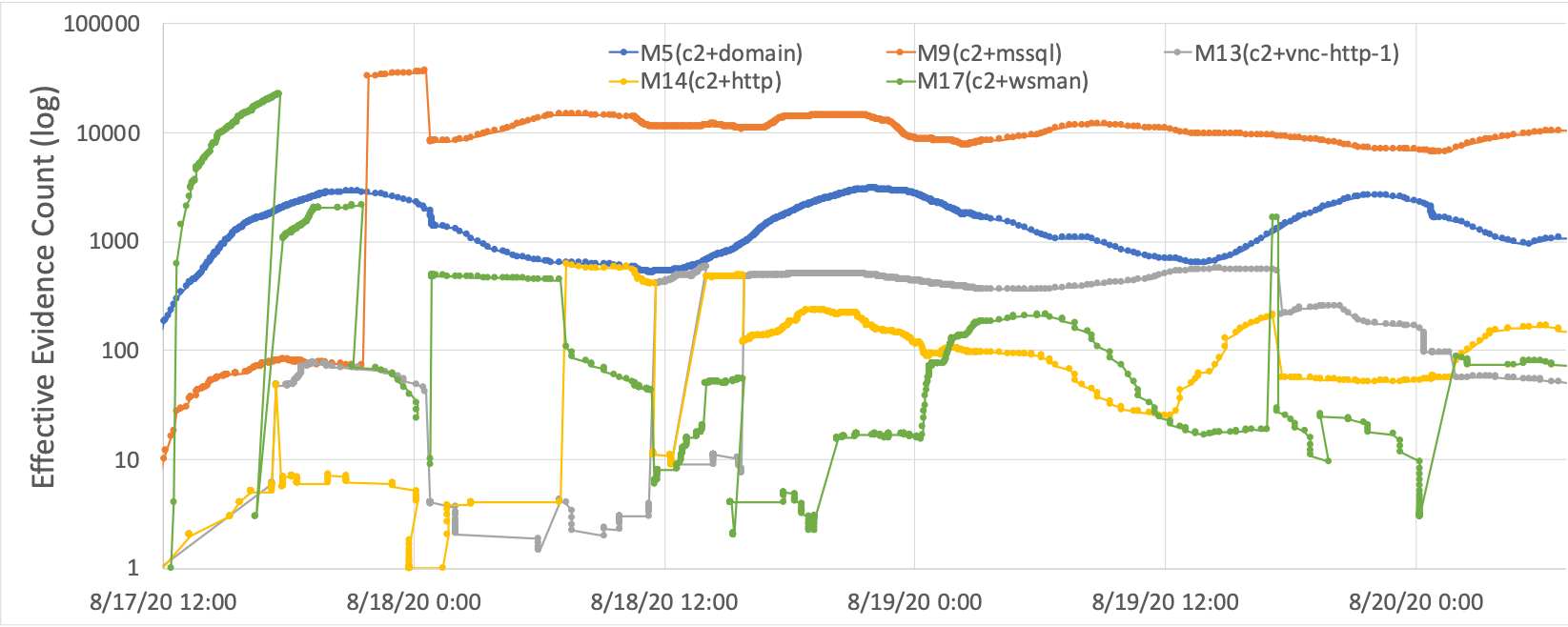}
    \caption{The effective evidence counts (alert volume in the moving window) of five models that have C2 as their characteristic feature for the attack stage component.}
    \label{fig:model3}
\end{figure*}

\subsection{C2 Activities through different services}
\label{sec:c2}
Through discussion with SOC analysts, C2 activities are of strong interests as they are indicative of potential persistent threats. Early detection of such can help prevent catastrophic cyber incidents. Therefore, we specifically look for attack models that have characteristic features of C2. Figure \ref{fig:model3} shows the `effective evidence counts' for five attack models that all have C2 as their characteristic feature in the attack stage component. The effective evidence count of a model is the EWMA of the alert volume within the moving window. This result reflects data collected between August 17 and August 20, 2020, around the time the higher education campus opened for the fall semester. 

Note the semi-periodicity of effective evidence counts for M5 (C2+Domain) shown in Fig. \ref{fig:model3} (second line from the top). This is the same model discussed in Section \ref{sec:c2+domain}, and confirms the continuous semi-periodic C2 activities through domain controllers. Additionally, we see C2 activities through other services. Among them, C2 through WSMAN (Windows Remote Management) was also observed earlier and C2 through HTTP related services are not surprising, either. Again, these may be truly malicious or due to system misconfigurations. An interesting observation here is the rise of C2 through MS-SQL, a type of activity not prevalent before mid August. Using ASSERT, SOC analysts would see this Model 8 emerging and persistently stay high volume (one or more orders or magnitude than other C2 activities). This could be due to an introduction of a service as the fall semester begins, which again could be just system misconfiguration or true malicious activities taking advantage of the new service.

\subsection{Summary of Usages and Benefits for ASSERT}
\label{sec:usessummary}
Over the course of collaborating with OmniSOC to deploy and test ASSERT, we identify many use-cases and benefits for ASSERT in SOC operations. An independent technical report summarizing the findings in deploying ASSERT in a SOC environment was produced by Kiser \etal \cite{Kiser2020}. With the use-cases described in this paper and additional experiences, we summarize the benefits of using ASSERT below.
\begin{itemize}
\item Reduce the reliance to interpret and analyze individual intrusion alerts by presenting a holistic view of ongoing and emerging attack behaviors.
\item Bring analyst's focus to the few critical emerging attack behaviors with their characteristic features, instead of burying in the process of sorting through overwhelming alerts. 
\item Track semi-regular events to help patch system vulnerabilities, improve IDS configurations, and/or identify persistent threats.
\item Analyze attack behaviors across organizations (by sharing anonymized characteristic features) could bring predictive and actionable threat intelligence.
\end{itemize}

\section{Conclusion}
\label{sec:conclusion}
This work builds on our previous research and develops a deployable, efficient, usable, and impactful ASSERT system. We describe the advances made to achieve near real-time unsupervised, continual learning of attack models based on information theoretic metrics. The advances are theoretically sound and practically efficient. Our deployment experiences and results have shown significant improvement in computational efficiency (improving from processing $\sim$20K alerts per hour to 100K+ using a standard server), the ability to summarize hundreds of thousands of alert into tens of contextually meaningful attack models with characteristic features, and, most importantly, bringing analysts attention quickly to critical attack behaviors that could take a long time to discover and sometimes overlooked. The research team is finalizing the development and will share ASSERT via a public repository.


\bibliographystyle{IEEEtran}
\bibliography{ref_2020}
\end{document}